\documentclass{iopjournal}

\usepackage[colorinlistoftodos]{todonotes}
\usepackage{natbib}
\usepackage{upgreek}
\usepackage{changepage}
\usepackage{amsmath}
\usepackage[table]{xcolor}
\renewcommand{\tabular}{\fontsize{8}{10}\selectfont \oldtabular}
\usepackage{float}
\usepackage{lineno}

\begin{document}


\title{Phase-contrast micro-CT for intra-operative breast tumour margin assessment using a microfocus x-ray source and photon-counting detector}

\author{Michelle K. Croughan$^{1*}$\orcid{0000-0001-9345-2345},
Timur E. Gureyev$^2$\orcid{0000-0002-1103-0649},
Jane Fox$^3$,
Mikkaela McCormack$^{2}$,
James A. Pollock$^{1}$\orcid{0000-0003-2594-9653},
Dominic Jurkschat$^1$\orcid{0009-0008-9515-1405},
Stephanie A. Harker$^{4}$,
and Marcus J. Kitchen$^{1}$\orcid{0000-0002-0029-6660}}

\affil{$^1$School of Physics and Astronomy, Monash University, Melbourne, Australia}

\affil{$^2$School of Physics, University of Melbourne, Melbourne, Australia}

\affil{$^3$Monash Health, Melbourne, Australia}

\affil{$^4$Peninsula University Hospital, Bayside Health, Melbourne, Australia}

\affil{$^*$Author to whom any correspondence should be addressed.}

\email{michelle.croughan@monash.edu}


\begin{abstract}
\textbf{Objective:} Intra-operative tumour margin assessment during breast-conserving surgery requires rapid, high-resolution imaging of excised tissue, allowing the surgical team to take appropriate action within a single operation. This study evaluates a custom propagation-based phase-contrast micro-computed tomography (micro-CT) system designed to meet these clinical constraints without specialised optical elements.

\textbf{Methods:} The experimental setup pairs a microfocus x-ray source with a photon-counting detector in a cone-beam geometry. We explore how the spatial coherence of the source can provide propagation-based phase contrast -- with no additional specialised optical elements -- and balance this against maximising the x-ray flux of the cone-beam geometry. System performance was evaluated across two anode target materials and filtration configurations at various tube power settings. Imaging capabilities were validated using anthropomorphic breast tissue phantoms and a formalin-fixed paraffin-embedded (FFPE) breast tissue specimen, with reconstructions compared against gold-standard histology.

\textbf{Results:} An unfiltered tungsten target operated at 40 kVp yielded optimal image quality. The optimised system achieved high-resolution CT reconstructions of a 5 cm diameter sample with an isotropic voxel size of 40.7 $\upmu\text{m}$ in a scan time of 12 minutes. Reconstructed volumes demonstrated strong visual correlation with corresponding histology slides.

\textbf{Conclusion:} Combining a microfocus source with a photon-counting detector enables high-resolution, phase-contrast micro-CT within a clinically viable timeframe, demonstrating strong potential for intra-operative margin assessment.

\end{abstract}

\section{Introduction}
\label{sec:introduction}
Breast-conserving surgery is an intervention which makes up 65.5\% of surgeries related to breast cancer in Australia and New Zealand~\citep{BQA_2022}. The excised breast tissue is examined to ensure the entire tumour has been removed by assessing whether the margins of the tissue contain part of the tumour (positive margin) or are tumour-free (negative margin). Margin assessment can either occur intra-operatively or post-operatively~\citep{luoRecentAdvancesIntraoperative2022}. If the \textbf{post}-operative margin assessment is positive, the patient may require a second operation. This can have negative impacts on the patient and health-care system. This paper focuses on \textbf{intra}-operative margin assessment, which can provide margin information quickly, allowing the surgical team to take appropriate action within a single operation. As intra-operative margin assessment techniques improve in terms of sensitivity and specificity, key outcomes will be reached, such as a reduction in secondary operations and a reduction in excessive tissue removal.

A suite of very different techniques exists for intra-operative assessment. Review articles are available that detail the current state of the art~\citep{maloneyReviewMethodsIntraoperative2018,pradiptaEmergingTechnologiesRealTime2020,luoRecentAdvancesIntraoperative2022,liDevelopmentIntraoperativeAssessment2022} and give meta analysis on their diagnostic accuracy \citep{stjohnDiagnosticAccuracyIntraoperative2017,dowlingDiagnosticAccuracyIntraoperative2024}. 
Of these, x-ray imaging is a promising technology that we focus on in this paper. Conventional 2D (specimen radiography, mammography) or 3D (tomosynthesis, cone-beam computed tomography) absorption-based x-ray images typically do not have clear contrast between different types of soft tissues. As such, research is being conducted globally to explore the potential for phase-contrast x-ray imaging to improve image quality for the application of intra-operative margin assessment, diagnostic imaging, and virtual histology, as detailed in the following paragraphs. Intra-operative 3D x-ray imaging of tissue specimens can have additional post-operative benefits beyond margin assessment. The images collected intra-operatively could also be used for precise localisation of the suspect tissue, guiding tissue block selection and processing in the pathology laboratory, potentially saving on storage, processing costs, and pathologists time. This dual purpose may make intra-operative 3D x-ray imaging particularly attractive in a clinical setting.

Shifting from 2D to 3D imaging techniques such as digital breast tomosynthesis and conventional micro-CT has been demonstrated to improve margin identification \citep{uranoDigitalMammographyDigital2016,parkDigitalBreastTomosynthesis2019,partainDifferencesReexcisionRates2020,kulkarniHighResolutionFull3DSpecimen2021,manhoobiDiagnosticAccuracyRadiography2022,streeterEmergingFutureUse2022a}. Micro-CT has been explored extensively in the literature using commercial absorption-contrast devices~\citep{tangMicrocomputedTomographyMicroCT2013,tangPilotStudyEvaluating2013,tangIntraoperativeMicrocomputedTomography2016,mcclatchyCalibrationAnalysisMultimodal2017,mcclatchyMicrocomputedTomographyEnables2018,qiuMicrocomputedTomographyMicroCT2018,janssenFeasibilityMicroComputed2019,dicorpoRoleMicroCTImaging2020,abelCanMicrocomputedTomography2020,streeterBreastConservingSurgeryMargin2023}. The signal-to-noise ratio (SNR), and contrast-to-noise ratio (CNR, see Sec.~\ref{sec: image quality}) between soft tissues in CT can be improved by also considering the phase information of the x-ray wavefield. Phase-contrast imaging and the subsequent phase-retrieval post-processing step are powerful tools for breast cancer imaging. The high coherence and parallel beam of synchrotron sources have made them ideal for developing phase-contrast imaging for 3D virtual histology~\citep{takedaPhasecontrastXrayCT1998,BaranAdjunctToHistopathology2018,aranapenaMultiscaleXrayPhasecontrast2023,donatoAdvancingBreastCancer2024a,donatoIntegratingXrayPhasecontrast2024b} and breast cancer diagnostic imaging~\citep{arfelli_low-dose_1998,gureyevPropagationbasedXrayPhasecontrast2019a,brombalMonochromaticBreastComputed2019a,longoAdvancementsImplementationClinical2019,tavakolitabaPropagationBasedPhaseContrastCT2021a}. Typical clinical sources do not produce a spatially coherent x-ray wavefield like that at a synchrotron facility, which is required for phase-contrast x-ray imaging. Fortunately, advancements in hardware and novel experimental techniques are allowing for the translation of phase-contrast imaging to x-ray imaging setups that is compact enough to be used clinically~\citep{streeterEmergingFutureUse2022a,donatoAdvancingBreastCancer2024a}. 

A prominent phase-contrast technique is edge-illumination~\citep{olivoCodedapertureTechniqueAllowing2007b}. This technique separates an incoherent x-ray wavefield into many small spatially coherent beamlets, making phase-contrast imaging possible with a clinical source. 
Edge-illumination CT has shown potential for intra-operative margin assessment~\citep{havariyounCompactSystemIntraoperative2019} with demonstrations of CT scan times of 10 minutes for 3 cm~\citep{massimiLaboratorybasedXrayPhase2019a,massimiVolumetricHighResolutionXRay2022b} and 5 cm~\citep{massimiDetectionInvolvedMargins2021c} sized samples. Edge illumination is experimentally advanced compared to traditional absorption-based micro-CT due to the introduction of specialised optical elements. Here, the significant improvement in image quality with the phase contrast must be weighed against the added cost and complexity of optical elements that must remain stable during imaging.

Microfocus x-ray sources produce a spatially coherent x-ray wavefield, allowing for propagation-based phase-contrast x-ray imaging; a technique requiring no additional optical elements and simply needs the x-ray wavefield to propagate after the sample, which is achieved by placing the detector some distance away from the sample~\citep{wilkins_phase-contrast_1996}. This technique is used for virtual histology to produce high-resolution 3D images of paraffin-embedded tissue~\citep{twengstromCanLaboratoryXray2022} with scan times of 100 to 180 minutes for 3 cm sized samples, and acetone fixed tissue~\citep{romell_x-ray_2026} with 2 to 3~hour scans for 2.5~cm samples. In order to be suitable for intra-operative imaging scan times can be reduced by having coarser spatial resolution and shorter propagation distances.
Given that propagation-based phase contrast does not rely on any special optical elements it may prove easier for clinical adoption from a manufacturing, cost and training perspective. As such, we believe it is valuable to explore propagation-based phase-contrast x-ray imaging as a potential tool for intra-operative tumour margin assessment. 

A microfocus x-ray source produces a polychromatic cone beam, which creates challenges for optimising the imaging setup. Filtering the x-ray beam and increasing the effective propagation distance to increase phase contrast reduces the x-ray flux at the detector, which in-turn increases image noise.
Thus, unlike with synchrotron radiation increasing the propagation distance in a cone-beam setup, to maximise the benefits of phase contrast, must be balanced with the concurrent loss in x-ray flux to ensure the best image quality.
Intra-operative tumour margin assessment must be fast to be a viable tool. Because the tissue is removed from the patient, radiation dose due to x-ray imaging is not important. Thus, the scan time is the main constraint for this application and the setup that maximises flux, at the appropriate x-ray energy range, will provide the highest quality images in the shortest time~\citep{nesterets_optimisation_2018}.

In this paper, we present a custom phase-contrast micro-CT setup using a microfocus x-ray source and a photon-counting detector. 
We explore how phase contrast resulting from the high spatial coherence of the microfocus source allows for noise reduction via phase retrieval while maximising the x-ray flux in the cone beam geometry.
A 12 minute scan time was deemed appropriate for intra-operative imaging. With our current setup, we can produce 3D images of 5 cm sized samples with a 40.7 $\upmu$m effective pixel size. We show tomograms of anthropomorphic breast tissue phantoms and formalin-fixed paraffin-embedded (FFPE) breast tissue as proof of concept. Propagation-based phase contrast can be used to improve image quality, but we also show that the setup should not be optimised for maximum phase contrast, as the loss of x-ray flux due to the cone-beam geometry counteracts the benefit of the additional phase contrast. This proof of concept demonstrates this technology as a potentially viable intra-operative margin assessment tool.

\section{Experimental details}
\begin{figure}
\centerline{\includegraphics[scale=1]{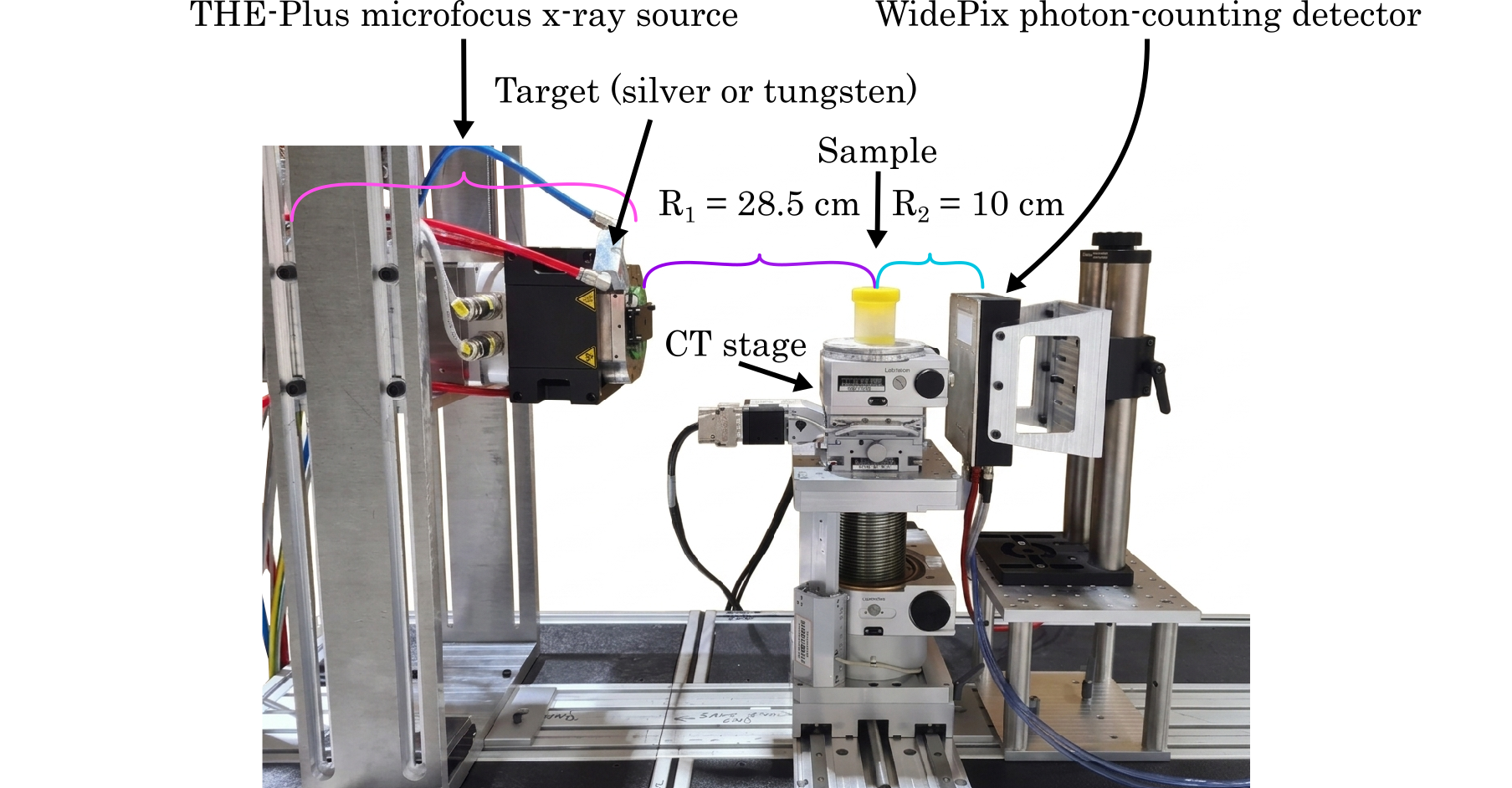}}
\caption{Photograph of the custom phase-contrast micro-CT setup, with microfocus x-ray source and photon-counting detector. }
\label{setup}
\end{figure}

The micro-CT setup used is shown in Fig.~\ref{setup}. This setup utilises advancements in x-ray imaging hardware to allow for fast acquisition of high-resolution 3D volumes. Details of the setup, sample preparation, and scan parameters are described in the following subsections.
\subsection{Microfocus x-ray source}
We employed the THE-Plus microfocus cone-beam x-ray source by \textit{X-RAY WorX GmbH,} Germany, which has a spot size with full width at half maximum (FWHM) on the order of microns that varies with the tube power and voltage settings. The small spot size means the x-ray wavefield is sufficiently spatially coherent for propagation-based phase contrast imaging, even with a polychromatic beam~\citep{wilkins_phase-contrast_1996} and a short propagation distance. Transmission style targets are used, enabling a small and uniform focal spot size. The target has a diamond substrate on the exit surface and a liquid cooled tube for high stability of the spot focus and beam intensity. This source has previously been demonstrated to be suitable for phase-contrast imaging~\citep{alloo_recovering_2023}. The x-ray spectrum must be optimised to differentiate between the different soft tissues within breast tissue: adipose, glandular, and tumour. We explored both silver and tungsten targets, with various filters: aluminium, silver, and tin, and different peak voltage (kVp) settings.

A silver target material produces characteristic x-rays with a dominant K-alpha peak at 22.16~keV that is well-suited for imaging soft tissue samples about 5 cm in size. Due to the relatively low melting point of silver ($962\,^\circ\text{C}$), the target power was kept at 10 W to prevent melting or needing to defocus the electron beam significantly at higher power settings. For the 6 $\upmu$m thick silver target, the maximum tube voltage is 50 kVp, else the electron beam will penetrate the target and result in lower x-ray flux. The tungsten target material produces characteristic x-rays with energies either too low or too high for the purpose of breast tissue imaging, leaving only bremsstrahlung radiation. The higher melting point of tungsten ($3422\,^\circ\text{C}$) allowed us to reach the maximum target power of 50 W. Increasing the target power also increases the focal spot size; however, varying the target power from 10 W to 50 W only worsened the spatial resolution by 1 $\upmu$m due to the low magnification used (Sec~\ref{subsec: geometric setup}), and so lower power settings were not explored. With the tungsten target, a tube voltage range of 40 kVp to 80 kVp was explore. A minimum of 40~kVp was tested as power limitations at kVp settings below 40 kVp resulted in a significant reduction in x-ray flux.

\subsection{Photon-counting x-ray detector}
We employed a WidePIX L $2\times5$ Medipix3 photon-counting detector by \textit{Advacam s.r.o.}, Czech Republic, to capture high-resolution images with minimal detector blurring by operating in charge-summing mode (CSM). The 10 sensors are arranged in a $2\times5$ array that has an area of $512\times1280$ pixels with a native pixel size of 55 $\upmu$m. The detector point spread function (PSF) was computed, by measuring the edge-spread function of a steel blade, to have a FWHM of approximately 1 pixel, with some variation based on the x-ray spectrum. The detector has a 1 mm thick cadmium telluride (CdTe) sensor, and was used in continuous mode with 12-bit depth. Threshold 0 was set to 5.9 keV and threshold 1 was set to 12.2 keV for CSM to reject photons below 12.2~keV. This setting helped minimise beam hardening without significantly limiting the flux penetrating through the sample.

\subsection{Geometry of the setup}
\label{subsec: geometric setup}
The geometry of the setup was constrained by needing to image samples up to 5 cm in size within the field of view of the WidePIX detector (7.1 cm wide with no magnification). Thus the maximum magnification factor $M$ possible would be 1.42. The magnification range of $M = 1$ to $M = 1.42$ will give an effective pixel size range of 55 to 38.7 $\upmu$m. Since the spot size of the microfocus source is much smaller than these possible effective pixel sizes the spatial resolution of the system is primarily limited by the effective pixel size~\citep{bidolaOptimizationPropagationbasedPhasecontrast2015a,lioliouFrameworkOptimizeFixedlength2024}. Thus, the largest magnification in this range provides the best spatial resolution. We used $M = 1.35$ to ensure the sample remained within the field of view for the whole scan, as the sample was manually positioned on the centre of the CT stage. This gave an effective pixel size of 40.7 $\upmu$m. 

The geometry determines the effective propagation distance $R' = R_1R_2/(R_1 + R_2)$, where $R_1$ is the source to sample distance and $R_2$ is the sample to detector distance. $R_1$ and $R_2$ can be increased or decreased while maintaining a constant magnification factor, as $M = (R_1 + R_2) / R_1$. Increasing the effective propagation distance increases the boost in SNR and CNR provided by phase retrieval. Conversely, due to the cone beam, this will also result in a decrease in x-ray flux onto the detector, which will increase the image noise. \cite{nesterets_optimisation_2018} explains, that under the constraint of a fixed image acquisition time, the geometry that maximises the x-ray flux of the system should provide the best image quality. This was validated by adjusting the intrinsic image quality metric~\citep{gureyev_signal--noise_2026} to account for the scan time rather than the radiation dose, showing that in general the closer the sample is positioned to the source, the better. Thus, $R_1$ and $R_2$ were set to 28.5 cm and 10 cm, respectively, to allow the most compact setup possible with $M=1.35$ and without the sample stage colliding with the detector, see Fig.~\ref{setup}. This compact setup was compared to one with a longer effective propagation distance in Sec.~\ref{sec: results}, specifically Fig.~\ref{propagation comparison}.

\subsection{Anthropomorphic breast tissue phantom and Formalin-fixed-paraffin-embedded breast tissue}
Anthropomorphic breast tissue phantoms were created using animal lard, egg white, and crushed egg shells. These materials have similar x-ray properties to tissues of the breast, representing adipose tissue, glandular tissue, and micro-calcifications, respectively~\citep{yakabeEffectDoseReduction2010,freedXrayPropertiesAnthropomorphic2011a,ikejimbaNovelPhysicalAnthropomorphic2017}. 
The phantoms were created by warming the lard to a viscous liquid and mixing it by hand with room-temperature egg whites. Micro-calcifications of different sizes were created by crushing eggshells and sieving them using steel geological sieves with aperture sizes of 53 and 180 $\upmu$m. Micro-calcifications with sizes ranging $53-180$ $\upmu$m were incorporated into the breast phantom. Water evaporated from the egg white over time, and so a new phantom was created for each set of image testing, leading to three different phantoms shown in Figs.~\ref{silver_recons},~\ref{tungsten_recons}, and~\ref{propagation comparison}.

An FFPE block of breast tissue was imaged to demonstrate that real tissues can be visualised effectively with this custom setup. The FFPE block was a $4 \times 2.8$~cm rectangle, with a maximum sample thickness of 4.88 cm. The sample contained ductal carcinoma in situ (DCIS) with micro-calcifications. CT slices are shown in Fig.~\ref{FFPE and hist} alongside a microscope image of a haematoxylin and eosin stained tissue slide taken directly from the top surface of the FFPE block.

\subsection{CT scan parameters}
Each 360-degree rotation scan was captured within 12 minutes. Nominally, 3718 projections were captured with a 193 ms exposure time per projection. However, in some configurations the exposure time needed to be reduced due to saturation of the detector, in which case the number of projections was increased to maintain the same total exposure time in each scan. Additionally, 50 flat-field images were collected before the scan to correct for spatial fluctuations in the beam intensity and pixel response.

\section{Image reconstruction and quality metrics}\label{sec: image quality}
The projections for each scan were flat-field corrected and ring correction was applied using Vo's all-stripe method~\citep{vo_superior_2018}. Phase retrieval using the Generalised Paganin Method~\citep{paganin_simultaneous_2002,paganin_boosting_2020,pollockPrecisePhaseRetrieval2022} was applied to the projections using a phase retrieval parameter, $\gamma'$, which has been adjusted to account for the spatial resolution of the imaging system~\citep{beltran_phase-and-amplitude_2018,gureyev_optimisation_2026}. Typically, one uses $\gamma = \delta / \beta$ in this phase retrieval method, where $\delta$ and $\beta$ are, respectively, the real and imaginary parts of the refractive index decrement. Phase retrieval smooths images, however the imaging system already applies some smoothing due to the finite spatial resolution, and so this is taken into account by computing $\gamma' = \gamma - 4 \pi \sigma_{sys}^2 / (2 R' \bar{\lambda})$ \citep{gureyev_optimisation_2026}, where $\sigma_{sys}$ is the standard deviation of the PSF of the imaging system and $\bar{\lambda}$ is the mean x-ray energy.
The $\gamma'$ values used varied based on power, kVp, target, and filter options for the scan, which are included in Table~\ref{tab:image quality}. 
Tomographic reconstruction was performed using the cone-beam FDK algorithm~\citep{feldkampPracticalConebeamAlgorithm1984} as implemented in the ASTRA toolbox~\citep{vanaarleASTRAToolboxPlatform2015,aarleFastFlexibleXray2016}, providing tomographic slices of the linear attenuation coefficient $\mu$ throughout the sample. The mean value of $\mu$ for the lard and egg white were used to estimate the mean x-ray energy $\bar{\lambda}$ using the xraylib package~\citep{schoonjans_xraylib_2011}.

The quality of the 3D images was determined by using four different metrics measured directly from the final reconstructed 3D image volume: 

1) The SNR of a region of egg white. As this is the more attenuating soft tissue it will have the most noise in the reconstruction, and so is a good indicator of the lowest SNR within the reconstruction. The SNR is computed by dividing the mean value of a region by the standard deviation in that region. The higher the SNR the less relative noise in the tomogram.

2) The CNR between a region of egg white and lard. This determines how easy it is to differentiate between two materials by taking into account the contrast between them and the level of noise in the tomogram. The difference between the mean values of these two materials is computed and divided by the standard deviation of one of the materials, whichever is greater. A larger value for CNR indicates better contrast between the two tissue types; increased CNR improves the likelihood of differentiating between very similar tissues such as glandular/benign fibrotic tissue and tumour tissue.

3) Spatial resolution is measured by fitting an edge profile using the Fileswell package~\citep{ahlers_high-energy_2025} to the boundary between the egg white and lard. This boundary allows us to determine the extent to which the interface between the two tissues types is resolved.
A Gaussian function is fit to the derivative of the edge profile to determine the standard deviation $\sigma$ of the PSF for the tomogram. The resolution is characterised as the FWHM of the Gaussian, where FWHM$ = 2 \sigma\sqrt{2\ln (2)}$. This measure of the spatial resolution for the reconstructed CT images was chosen as it directly measures the boundary width between the two primary materials of interest.

4) It is important to consider both spatial resolution (FWHM) and CNR together as often changing one setup parameter can affect both. For example, increasing the x-ray source power will increase the x-ray flux, improving CNR, but also result in a larger focal spot size making the spatial resolution more coarse. The spatial resolution and CNR can be combined to create a metric in a similar manner to the biomedical x-ray imaging quality characteristic~\citep{gureyev_signal--noise_2025}. In this case we do not need to account for the radiation dose, seeking only the highest CNR achievable with the finest spatial resolution in the reconstructed volume. Thus, $\text{CNR} / \text{FWHM}^{3/2}$ provides a simple score to correctly balance these two metrics, and allows for direct comparison between different CT setups where both the spatial resolution and CNR vary. 

\section{Results}\label{sec: results}
\begin{figure}[H]
\centerline{\includegraphics[width=\columnwidth]{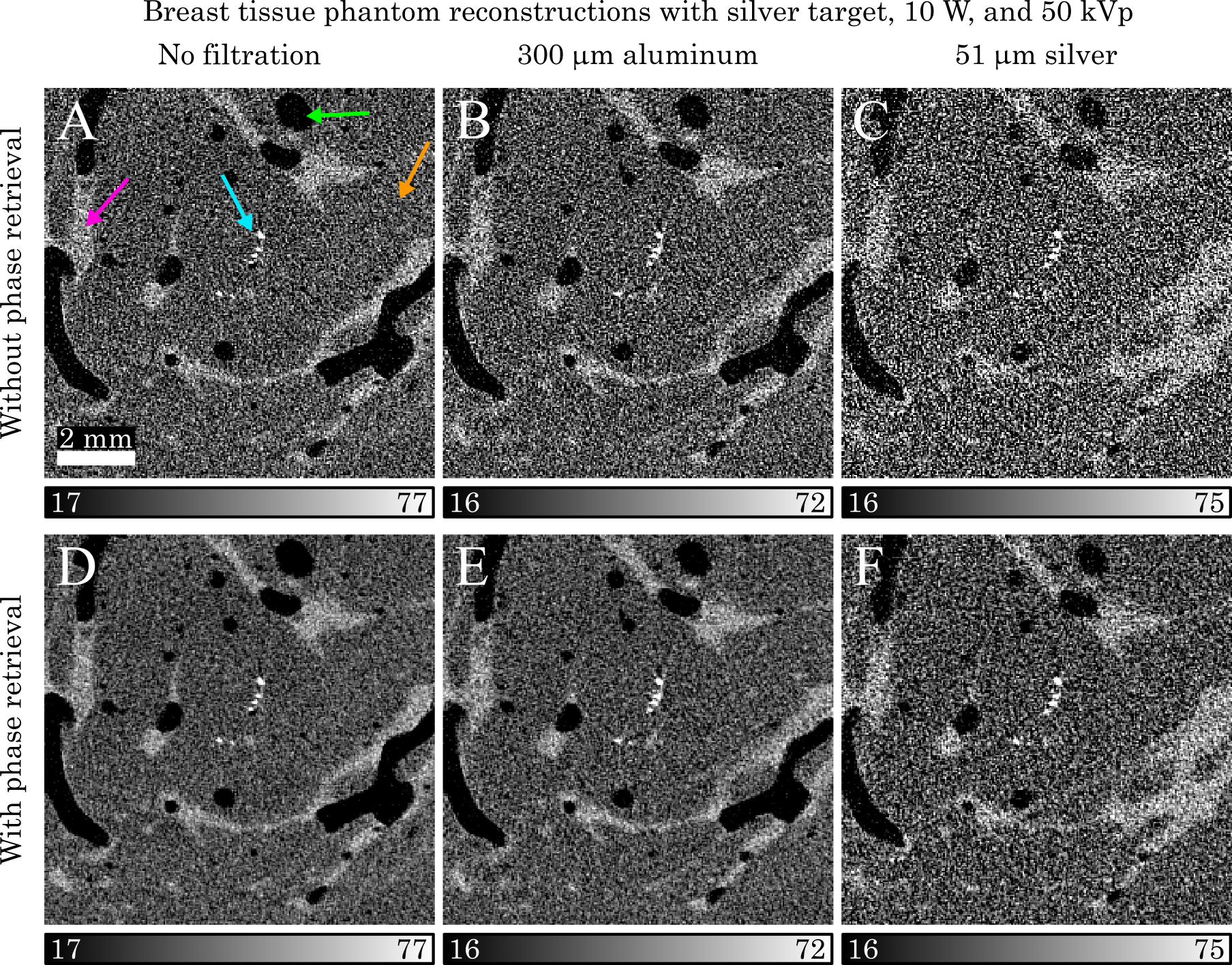}}
\caption{Tomographic slices of a breast tissue phantom, all captured using a silver target with 10 W and 50 kVp in a 12 minute scan time, with different filter options. A and D: no filter. B and E: 300~$\upmu$m of aluminium. C and F: 51~$\upmu$m of silver. The top row shows reconstruction without phase retrieval, and the bottom row shows with phase retrieval. Arrows indicate air bubbles (green), lard (orange), egg white (pink) and eggshell fragments (blue).}
\label{silver_recons}
\end{figure}

\begin{figure}[t]
\centerline{\includegraphics[width=\columnwidth]{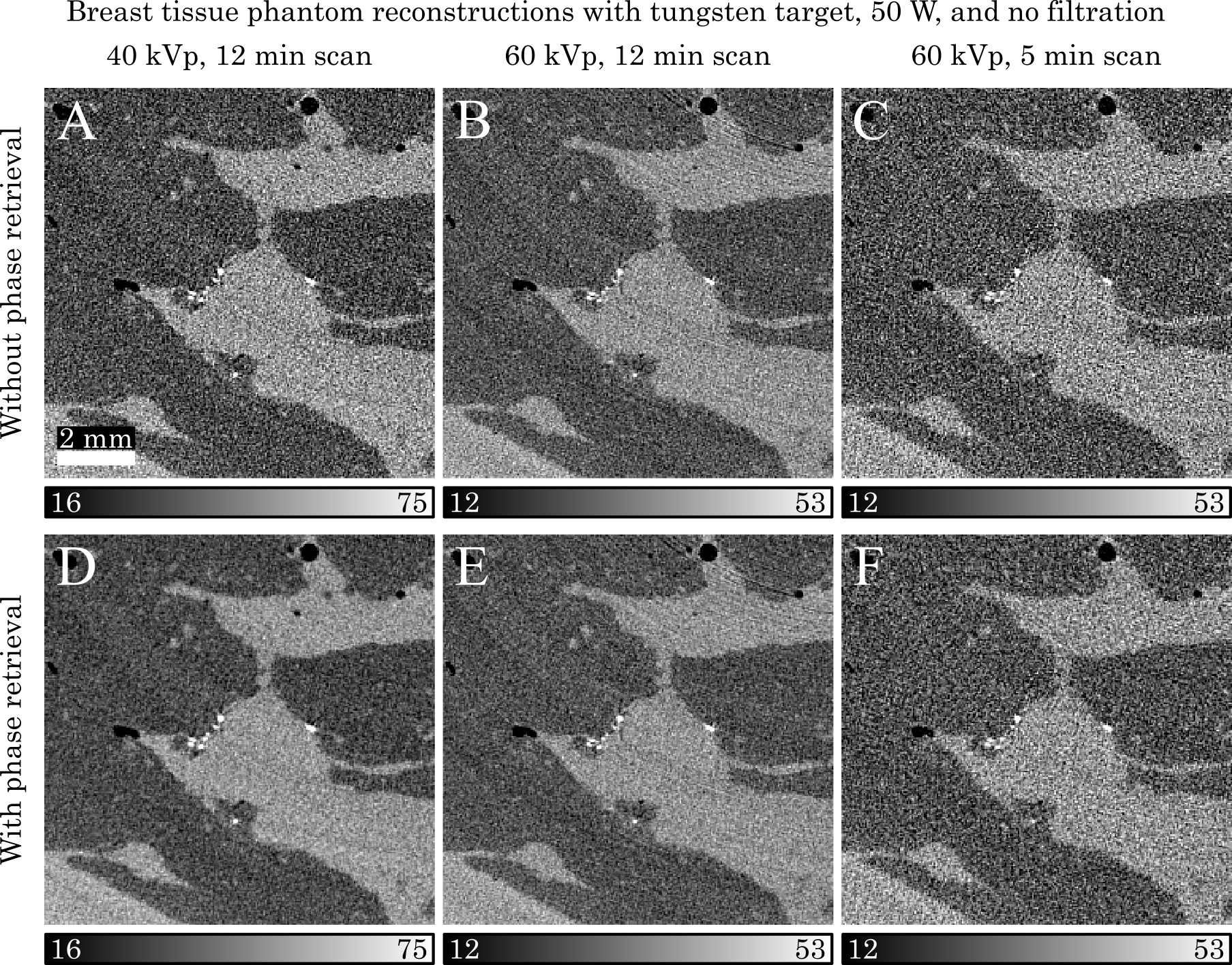}}
\caption{Tomographic slices of a breast tissue phantom, all captured using a tungsten target with 50 W in a 12 minute (A,B,D,E) or 5 minute (C,F) scan time. A and D: 40 kVp. B, C, E, and F: 60 kVp. The top row shows reconstruction without phase retrieval, and the bottom row shows with phase retrieval.}
\label{tungsten_recons}
\end{figure}

\begin{figure}[t]
\centerline{\includegraphics[width=\columnwidth]{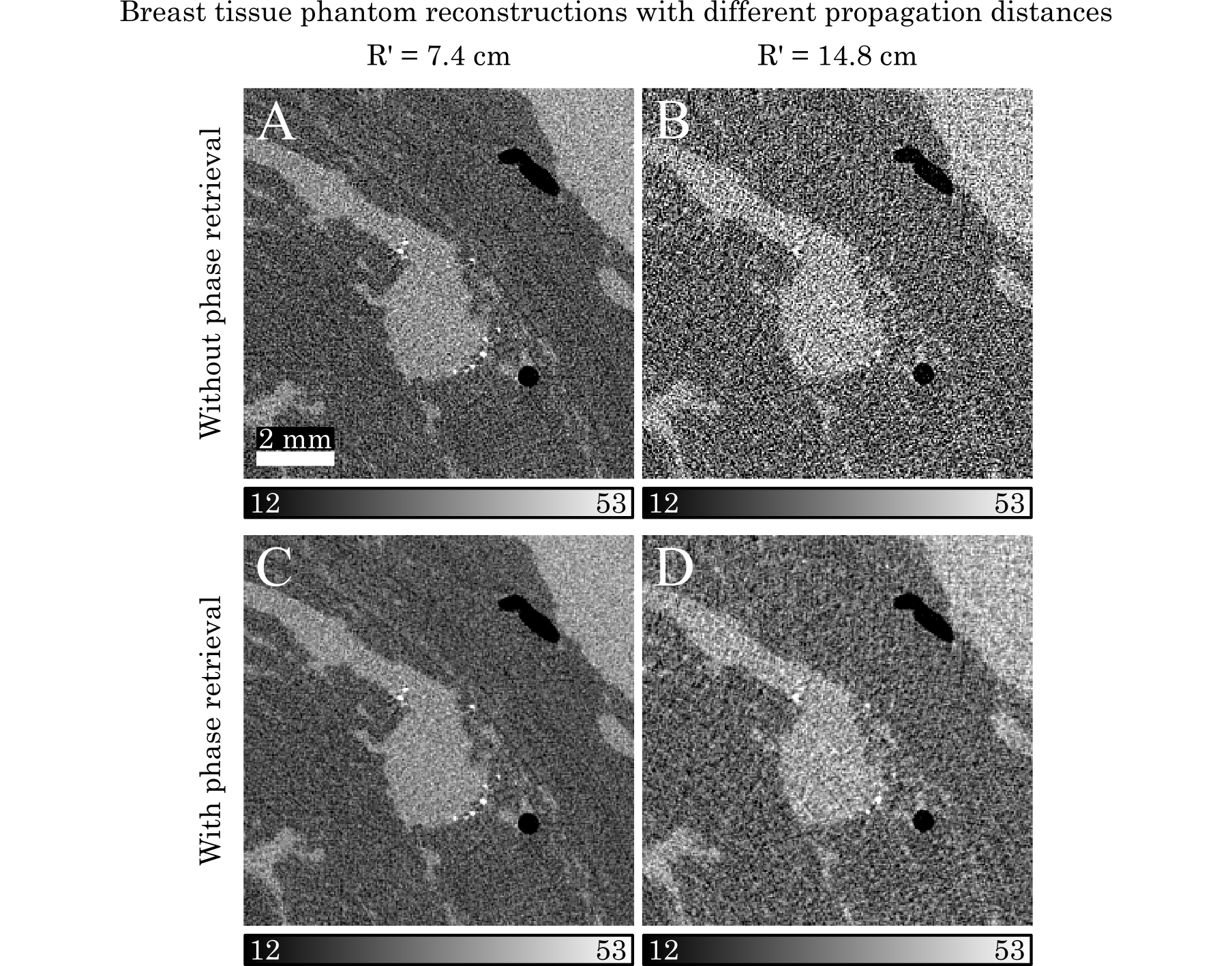}}
\caption{Tomographic slices of a breast tissue phantom, captured using a tungsten target with no filtration, 60 kVp, and 50 W in a 12 minute scan time. A and C: with an effective propagation distance $R'$ of 7.4 cm, B and D: with $R'$ of 14.8 cm. The top row shows reconstruction without phase retrieval, and the bottom row shows with phase retrieval.}
\label{propagation comparison}
\end{figure}

Different source settings, targets, and filter combinations were tested. Silver and tungsten target materials were used and filtration options were selected to create x-ray spectra that consisted mostly of energies between 16 kVp and 30 kVp, which is optimal for distinguishing between breast tissue types. CT scans with no filtration were also collected. Reconstructions are shown with and without phase retrieval applied, and in all images we can identify all three materials within the phantom. All scan parameters and the image quality metrics calculated using the reconstructions are shown in Table~\ref{tab:image quality}, followed by a discussion of key results. Different scan parameters resulted in different x-ray spectra, and the energy-dependent $\mu$ values for the different materials vary between the scans. To adjust for this, the x-ray images in Figs.~\ref{silver_recons}, \ref{tungsten_recons}, and~\ref{propagation comparison} have the gray scale range set to be from 0.5$\times \mu_{\text{lard}}$ to 1.5$\times \mu_{\text{egg white}}$. When imaging with 80 kVp the propagation-based phase contrast is negligible due to a larger fraction of higher energy photons, and so phase retrieval was not applied.

\definecolor{pastelpurple}{RGB}{230,220,250}
\definecolor{pastelgreen}{RGB}{180,240,190}

\begin{table}[t]
\caption{Comparison of image quality metrics with and without phase retrieval.}
\label{tab:image quality}
\centering
\setlength{\tabcolsep}{3pt}
\begin{tabular}{|c|c|c|l|c|c|c|c|c|c|c|c|c|}
\hline
\textbf{Target} &
\textbf{kVp} & \textbf{Power} &
\textbf{Filter} &
\multicolumn{4}{c|}{\textbf{No phase retrieval}} &
\multicolumn{5}{c|}{\textbf{With phase retrieval}} \\
\cline{5-13}
\textbf{material}& & (W) & ($\upmu$m) &
\textbf{CNR} & \textbf{SNR} & \textbf{FWHM} & \textbf{CNR/} & $\gamma'$ &\textbf{CNR} & \textbf{SNR} & \textbf{FWHM} & \textbf{CNR/}  \\
& & & & & & ($\upmu$m) & \textbf{FWHM$^{3/2}$} & & & & ($\upmu$m) & \textbf{FWHM$^{3/2}$}\\
\hline
\rowcolor{pastelpurple} Ag & 50 & 10 & None             & 1.25 & 3.83 & 46 & 4.1$e^{-3}$  & 370 & 1.56 & 4.82  & 62 & 3.2$e^{-3}$ \\
Ag & 50 & 10 & Al 300           & 1.13 & 3.43 & 58 & 2.6$e^{-3}$  & 396 & 1.88 & 5.79  & 72 & 3.1$e^{-3}$ \\
Ag & 50 & 10 & Ag 51            & 0.8  & 2.14 & *** & ***         & 392 & 1.36 & 3.65  & 79 & 1.9$e^{-3}$ \\
\rowcolor{pastelgreen} W  & 40 & 50 & None             & 2.33 & 6.32 & 64 & 4.5$e^{-3}$  & 356 & 3.77 & 10.31 & 81 & 5.2$e^{-3}$ \\
W  & 40 & 50 & Al 300           & 2.05 & 5.63 & 68 & 3.7$e^{-3}$  & 375 & 3.38 & 9.29  & 86 & 4.2$e^{-3}$ \\
W  & 60 & 50 & None             & 2.46 & 7.84 & 70 & 4.2$e^{-3}$  & 284 & 2.93 & 9.4  & 80 & 4.1$e^{-3}$ \\
W  & 60 & 50 & Ag 51            & 1.27 & 4.32 & 53 & 3.3$e^{-3}$  & 297 & 1.54 & 5.24  & 60 & 3.3$e^{-3}$ \\
W  & 60 & 50 & Sn 76            & 1.43 & 4.74 & 67 & 2.6$e^{-3}$  & 296 & 1.73 & 5.72  & 76 & 2.6$e^{-3}$ \\
*W & 60 & 50 & None        & 1.52 & 4.79 & 91 & 1.8$e^{-3}$  & 285 & 1.82 & 5.76 & 88 & 2.2$e^{-3}$ \\
**W & 60 & 50 & None & 1.16 & 3.86 & 64 & 2.2$e^{-3}$  & 245 & 2.11 & 7.03 & 83 & 2.8$e^{-3}$ \\
W  & 80 & 50 & None                 & 2.03 & 7.39 & 78 & 3.0$e^{-3}$  & - & -     & -     & -   & -\\
\hline

\end{tabular}
\newline
\newline
\fontsize{8}{10}\selectfont Ag = silver, W = tungsten, Al = aluminium, Sn = tin, CNR = contrast-to-noise ratio, SNR = signal-to-noise ratio, FWHM = full width at half maximum of egg/lard boundary edge profile, *5 minute scan time, **double effective propagation distance, *** reconstruction too noisy to compute. Purple highlight for finest spatial resolution scan, and green highlight for highest CNR and CNR/FWHM$^{3/2}$ scan.
\end{table}

Scans with the silver target (all 50 kVp and 10 W) and different filtration options are shown in Fig.~\ref{silver_recons}. An aluminium filter of 300~$\upmu$m was tested to filter low x-ray energies ($<$15~keV), as these would not improve image quality and result in beam hardening artefacts due to being mostly attenuated by the sample. However, the photon-counting detector has the built-in capability to ignore x-ray photons below a certain energy, which was set at 12.2 keV. It also weights all detected photon energies evenly, unlike integrating detectors that weight towards high energy photons~\citep{danielsson_photon-counting_2021}, which boosts the CNR. These effects, combined with the relatively low voltage, resulted in no beam hardening being observed in the scan with no filtration. In fact, the aluminium filtration resulted in worse image quality as it attenuated photons at all energies, resulting in lower SNR and CNR (see Table~\ref{tab:image quality}). A 51~$\upmu$m thick silver filter was also used to preferentially absorb photons with energies lower and higher than the characteristic peaks around 22~keV, making the x-ray beam more monochromatic. Despite this, it was found that filtering did not improve image quality due to the reduced x-ray flux (see Table~\ref{tab:image quality}).

Scans were captured using the tungsten target with peak voltage settings ranging from 40 kVp to 80 kVp. The higher source power of 50~W enabled higher x-ray flux, improving the CNR and SNR. Filtering of the beam with the tungsten target also resulted in worse image quality compared to no filtration. The best image quality was obtained with no filtration with 40 kVp and second best with 60 kVp, shown in Fig.~\ref{tungsten_recons} with metrics in Table~\ref{tab:image quality}. The 40 kVp tube voltage setting results in greater contrast between the lard and egg white due to the greater difference in the material's $\mu$ values, compared to higher kVp settings. An additional scan of the breast tissue phantom was collected with a reduced time of 5 minutes at 60 kVp, 50 W (Fig.~\ref{tungsten_recons} C and F). The CNR and SNR for the 5 minute scan both scored higher than the scans collected with the silver target (see Table~\ref{tab:image quality}). However, the spatial resolution for this scan was measured to be almost twice as coarse as the silver scans (see Table~\ref{tab:image quality}) due to the image noise limiting the measured spatial resolution.

As discussed in Sec.~\ref{subsec: geometric setup} the geometry that maximises the x-ray flux should provide the best image quality~\citep{nesterets_optimisation_2018}. A scan with double the effective propagation distance was collected to confirm this. The geometry was changed from $R_1 = 28.5$ cm and $R_2 = 10$ to $R_1 = 57$ cm and $R_2 = 20$ cm to give $R' = 14.8$ cm, keeping $M = 1.35$. This scan was captured with the tungsten target at 60~kVp, source power of 50~W, no filtration, and 12 minute scan time. Figure~\ref{propagation comparison} compares the reconstructions with $R'$ = 7.4~cm and 14.8~cm. The longer propagation distance resulted in lower flux and thus worse image quality (see Table~\ref{tab:image quality}) that was not compensated for by the increased propagation-based phase contrast.

The reconstruction with the best spatial resolution, highlighted in purple in Table~\ref{tab:image quality}, is the silver target scan with no filter and no phase retrieval (Fig.~\ref{silver_recons} A). The scan with the best CNR and SNR, highlighted in green in Table~\ref{tab:image quality}, was collected with the tungsten target at 40 kVp with phase retrieval (Fig.~\ref{tungsten_recons} B). The latter also received the best $ \text{CNR} / \text{FWHM}^{3/2}$ score. Phase retrieval improved $ \text{CNR} / \text{FWHM}^{3/2}$ for most scans, indicating overall better image quality. 

\begin{figure}[t]
\centerline{\includegraphics[width=\columnwidth]{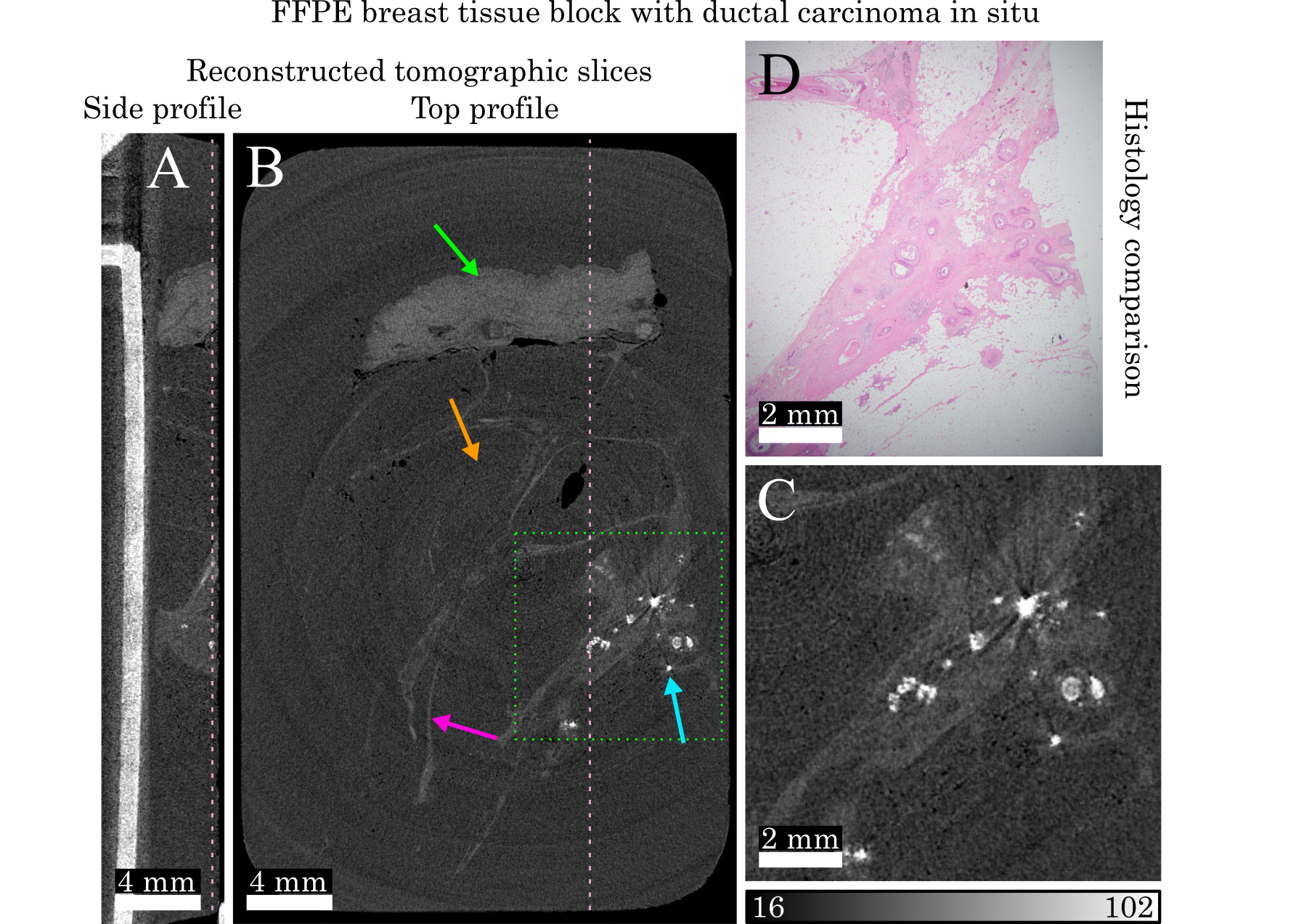}}
\caption{FFPE block of real breast tissue imaged with the optimum parameters for the custom phase-contrast micro-CT setup. A and B: Slices from the tomographic reconstruction showing a side (A) and top (B) profile, with pink dashed lines indicating relative locations. Arrows indicate skin (green), adipose tissue (orange), fibrotic tissue (pink) and micro-calcifications within ducts (blue). C: Zoomed in region of B highlighted with green dashed box. D: Visible light microscope image of histology slide showing similar region of tissue to that in C. X-ray images (A, B and C) all have the same gray scale.}
\label{FFPE and hist}
\end{figure}

The FFPE breast tissue block was imaged with the best setup parameters: tungsten target, 40~kVp, 50~W and no filter, as determined by the highest $\text{CNR} / \text{FWHM}^{3/2}$ score (see Table~\ref{tab:image quality}). This reconstruction is shown in Fig.~\ref{FFPE and hist} alongside a visible light microscope image of a haematoxylin and eosin stained tissue slide, used for histology, taken directly from the top surface of the FFPE block. Exact correlation between the slide and reconstruction was not expected, as the tissue slide was taken before x-ray imaging, and hence this tissue layer was not present in the FFPE block used in the x-ray imaging, but allows a comparative assessment of the level of detail shown in the x-ray CT.

In Fig.~\ref{FFPE and hist} that the adipose tissue has very similar x-ray properties to the paraffin wax, and so the boundary between the two materials is not visible in the CT reconstruction; however, for intra-operative imaging the tissue will not be paraffin embedded, and so the outer boundary of the tissue will be clear. Different tissue types such as skin, fibrotic/glandular tissue and micro-calcifications are clearly identifiable in the x-ray reconstruction. Similar to the histology slide, the x-ray reconstruction can clearly identify the boundary between adipose and glandular tissue. Visual correlation between the two imaging modalities is also seen in ducts containing ductal carcinoma in situ (DCIS) due to the presence of high contrast micro-calcifications. This level of detail in the intra-operative images could also assist pathologists in post-operative specimen assessment and histology tissue selection, where abnormalities can be difficult to identify with visual inspection alone (e.g. the location of DCIS within the tissue), providing dual purpose to the intra-operative imaging.

\section{Discussion}

X-ray micro-CT provides a mechanism to quickly assess cancer margins in intra-operative settings. In this paper, we presented an optimised phase-contrast micro-CT setup that uses both a microfocus x-ray source and a photon-counting detector, allowing for high-resolution imaging. With the current limitation of a 7.1~cm field of view for the detector, and 5~cm diameter sample size, the effective pixel size was 40.7~$\upmu$m. This allowed for the boundary between the materials representing glandular and adipose tissues to be resolved at well under 100~$\upmu$m resolution, and micro-calcifications with diameters ranging from 50 to 180~$\upmu$m were clearly observed.

Cone-beam geometry results in a complicated balance between propagation distance and other key scan parameters, unlike parallel beam geometry at a synchrotron where propagation-based phase contrast provides clear benefits. Drawing inspiration from previous optimisation work~\citep{nesterets_optimisation_2018,gureyev_signal--noise_2026} we demonstrated that with a fixed 12 minute scan the setup that maximised the x-ray flux at the optimal x-ray energies provided the best image quality. Strategies to optimise for phase contrast, such as filtering the x-ray beam or using larger effective propagation distances, resulted in worse image quality due to the associated reduction in x-ray flux for the scan duration. Despite optimising for flux, phase contrast and phase retrieval still provides benefits as it can improve CNR at a minimal expense of spatial resolution.

A range of imaging parameters were varied with image quality measured for each reconstruction, displayed in Table~\ref{tab:image quality}. The reconstruction with the best $\text{CNR} / \text{FWHM}^{3/2}$ score was collected with the tungsten target, no filtration, 40 kVp, 50 W and with phase retrieval applied. This reconstruction used a geometry of $R_1 = 28.5$ cm and $R_2 = 10$ cm which allowed for the maximum geometric magnification of $M=1.35$ limited by the size of the detector field of view. The spatial resolution of the egg white and lard boundary (representing the glandular and adipose tissue interface) was 81~$\upmu$m FWHM. We acknowledge that image quality metrics are a useful tool for quantifying and comparing images but are only part of the story when it comes to medical imaging. Radiologists and other medical image analysts preferences and training may affect the balance between CNR and spatial resolution that is considered ideal. As this research approaches clinical application this will be need to be taken into consideration.

Spatial resolution of intra-operative tumour margin assessment imaging systems, with under 15 minute scan times, varies for different absorption-contrast micro-CT and edge-illumination setups. For absorption-contrast, micro-CT reports of spatial resolution range from 150 $\upmu$m to 240 $\upmu$m ~\citep{mcclatchyCalibrationAnalysisMultimodal2017,streeterBreastConservingSurgeryMargin2023}. Edge illumination, an alternative imaging technique that also utilises phase contrast, has been conducted with a detector FWHM spatial resolution of 120 $\upmu$m~\citep{massimiLaboratorybasedXrayPhase2019a}. Our custom phase-contrast micro-CT setup produces 3D volumes with a voxel size of 40.7 $\upmu\text{m}^3$ and, for our optimum image quality, a spatial resolution FWHM of 81~$\upmu$m, which helps provide clear visualisation of tumour boundaries and other key features such as microcalcifications.

Breast tissue in the form of an FFPE block was imaged with the optimum imaging conditions for this custom setup, shown in Fig.~\ref{FFPE and hist}. This demonstrates that the results curated from the breast tissue phantoms translate well to imaging real tissue. Views from multiple directions can be seen, allowing for the assessment of margins from all angles, without overlapping tissue reducing clarity as with 2D projection images. Given this proof of concept work to produce high-resolution images, many future directions are possible. In the future, we plan to image excised breast tissue to assess the imaging capabilities with fresh tissue compared to the phantoms and FFPE tissue presented here, as well as address other challenges such as image artefacts from metal biopsy clips and tissue motion when not fixed. After that, we will investigate if this custom phase-contrast micro-CT setup can improve sensitivity and specificity of tumour margin assessment over current clinical intra-operative practices.

Improvements to the present micro-CT setup can be made in future. A higher powered microfocus x-ray source that can deliver more flux without increasing the spot size could be used to either provide an increase in CNR or reduce the scan time, without worsening the spatial resolution. A large detector could allow more geometrical magnification and thus better spatial resolution while still fitting the sample within the field of view. Finer steps in tube voltage can be tested to ensure an optimum unfiltered x-ray spectrum. Finally, geometrical magnification and the x-ray spectrum could be optimised for different sized samples, both smaller and larger than 5 cm. Beyond optimising the micro-CT septup for margin assessment, further research can be conducted into: how different pathological changes appear in the reconstructed image (for example: DCIS with and without calcification, duct sizes, and invasive or in situ tumour architecture), how adjusting imaging parameters can make particular features more or less prominent, and how this information can have clinical utility for post-operative analysis in the pathology laboratory.

%
%

\ack{The authors acknowledge Maria Chavez and Mayu Uemura for their useful discussion about the pathology of the FFPE breast tissue block and for providing microscope images of the histology slide shown in Fig.~\ref{FFPE and hist}. Additionally, we acknowledge Elette Engles for guidance on anthromothic breast tissue phantom preperation.}

\funding{Funded by Australian NHMRC Synergy Grant 2011204}
 This section is a list of funder names and grant numbers

\roles{\textbf{Michelle K. Croughan}: conceptualisation, data curation, formal analysis, investigation, methodoloy, project administration, resources, validation, writing - orginal draft, review, and editing. \textbf{Timur E. Gureyev}: conceptualisation, funding acquisition, methodology, project administration, supervision, writing - review and editing. \textbf{Jane Fox}: Conceptualisation, funding acquisition, writing - review and editing. \textbf{Mikkaela McCormack}: investigation, writing - review and editing. \textbf{James A. Pollock}: software, writing - review and editing. \textbf{Dominic Jurkschat}: software, writing - review and editing. \textbf{Stephanie A. Harker}: data curation, writing - review and editing. \textbf{Marcus J. Kitchen}: Conceptualisation, funding acquisition, methodology, project administration, resources, supervision, writing - review and editing.}

\data{Data is available upon resonable request. Imaging data of FFPE block tissues can not be made publically avaliable due to ethics restrictions.}

\bibliography{biblio}

\end{document}